\documentclass[12pt]{article}
\usepackage[latin1]{inputenc}
\usepackage{graphicx}
\usepackage{makeidx}
\usepackage{amssymb}
\usepackage{graphicx}

\usepackage{pdfpages} 

 \newcommand{\sinc}{{\rm sinc}}

 \newcommand{\kr}{k_{\rho}}

 \newcommand{\pa}{\partial}

 \newcommand{\text}{\rm}

 \newcommand{\ug}{ \; = \; }

 \newcommand{\bb}{\begin{equation}}
 \newcommand{\ee}{\end{equation}}
 \newcommand{\bc}{\begin{center}}
 \newcommand{\ec}{\end{center}}
 \newcommand{\bega}{\begin{eqnarray}}
 \newcommand{\ega}{\end{eqnarray}}
 \newcommand{\begae}{\begin{eqnarray*}}
 \newcommand{\egae}{\end{eqnarray*}}

 \newcommand{\h}{\hspace*{4ex}}
 \newcommand{\dis}{\displaystyle}

 \newcommand{\om}{\omega}

 \newcommand{\cent}{\centerline}
 \newcommand{\vs}{\vspace*}

\begin{document}
\baselineskip 0.5cm

\begin{center}

{\large {\bf Acoustic (Ultrasonic) Non-Diffracting Beams: Some theory, and
{\em Proposals of Acoustic Antennas} for several purposes $^{\: (\dag)}$ }} \footnotetext{ $^{\: (\dag)}$
E-mail addresses: \ recami@mi.infn.it ; \ mzamboni@decom.fee.unicamp.br }

\end{center}

\vs{5mm}

\cent{ Michel Zamboni-Rached$^{\; 1,2}$ \ and \ Erasmo Recami$^{\; 2,3,4}$ }

\vs{0.2 cm}

\centerline{$^{1}$ {\em Photonics Group, Electrical \& Computer Engineering, University of Toronto, CA}}
\centerline{$^{2}$ {\em DECOM, FEEC, Universidade Estadual de Campinas
(UNICAMP), Campinas, SP, Brazil}}
\cent{$^{3 \;}$ {\em Facolt\`a di Ingegneria, Universit\`a statale di
Bergamo, Bergamo, Italy.}}
\cent{$^{4 \;}$ {\em INFN---Sezione di Milano, Milan, Italy.}}

\vs{0.5 cm}


{\bf Abstract  \ --} \  On the basis of a suitable theoretical ground, we study and propose Antennas for the generation,
in Acoustics, of Non-Diffracting Beams of ultrasound. \ We consider for instance a frequency of about 40 kHz, and foresee
fair results even for finite apertures endowed with reasonable diameters (e.g., of 1 m), having in mind various possible applications, including remote sensing. \ We then discuss the production in lossy media of ultrasonic beams resisting both diffraction and attenuation. \ Everything is afterward investigated for the cases in which {\em high-power} acoustic transducers are needed (for instance, for detection at a distance ---or even explosion--- of buried objects, like Mines).

\

[Keywords: Acoustic Non-Diffracting Beams; Truncated Beams of Ultrasound; Remote sensing;
Diffraction, Attenuation, Annular transducers, Bessel beam superposition, High-power ultrasound emitters,
Beams resisting diffraction and attenuation, Acoustic Frozen Waves, Detection of buried objects, Explosion
of Mines at a distance].

\

\section{Introduction}

In this paper we aim at reporting about work performed by us during the last few years on theory
and generation (in Acoustics) of Non-Diffracting Beams[1,2]of ultrasound; having in mind various
possible applications, including remote sensing. \ In the first part of this paper, we shall
not deal, however, with the ``(Acoustic) Frozen Waves", confinig ourselves here to quote other articles,
like Refs.[3,4], in which they have been investigated.

\h Acoustic Non-Diffracting Waves (ANDW) were first studied, generated, and applied by Lu et al., starting
with 1992, for the particular, interesting case of the so-called (ultrasonic) X-shaped waves (see, e.g.,
Refs.[5,6]). \ For reviews about Non-Diffracting Waves (NDW), including X-shaped waves (as well as
Frozen Waves), one can see for instance Refs.[7,8] besides the initial Chapters in the already
quoted [1,2].

\h The NDWs (including of course the ANDWs) arose interest because of
their spatio-temporal localization, unidirectionality,
soliton-like nature, and self-healing properties[1,2]: All of them bearing interesting consequences,
from theoretical and experimental points of view, in all sectors of
physics in which a role is played by a wave equation.  The NDWs would keep such properties
all along an infinite distance, only in the ideal case implying an infinite energy flux through
any transverse plane. Such ideal NDWs cannot be practically generated, of course; and careful work
was needed for finding out analytic expressions for realistic NDWs --for example {\em truncated}---,
and then producing them (see Refs.[9,10] and refs. therein). Any realistic, finite-energy NDW will maintain
its good properties only within its depth of field: {\em much} longer, however, than the one reached by a
{\em diffracting} wave like the gaussian ones[1,2].

\h We are going to consider the problem of the truncated pulses in general (in electromagnetism, say), before
passing to Acoustics.

\subsection{{\em Analytic} Expressions for {Truncated}\\ Non-Diffracting Pulses} 

Let us go go on, therefore, to the problem of constructing in analytic form truncated Non-Diffracting Waves, in order
to be able to produce them experimentally. We address here the case of {\em pulses}, since the case of beams
have been extensively exploited elsewhere (see, e.g., Refs.[9,10] and refs. therein).

\h When one truncates an ideal non-diffracting pulse (INDP), the resulting wave field cannot be
obtained, in general, in analytic form. One has to resort, instead, to
the diffraction theory and perform numerical evaluations
of the diffraction integrals, such as that, well known, of
Rayleigh-Sommerfeld. \ And, indeed, one can get important pieces of information about
a truncated non-diffracting pulse (TNDP) by performing numerical
simulations of its longitudinal evolution, especially when
the pulse is axially symmetric.

\h However, let us mention first of all the possibility of obtaining truncated non-diffracting pulses
in analytic form by a heuristic method.\ Subsequently, we are going to show how the solutions
forwarded by our efficient method, expounded in Ref.[9] for beams, can be transformed into closed
form expressions for truncated non-diffracting pulses[11].

\subsection{A heuristic approach}

\begin{figure}[!h]
\begin{center}
 \scalebox{2.}{\includegraphics{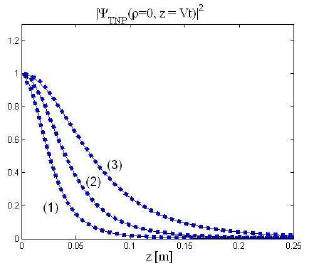}}
\end{center}
\caption{(Color online) Peak intensity evolution of a subluminal TNDP
for three choices[12] of the parameters \ [(1) \ $V=0.995c\;$ and $\;b=1.5\times
10^{15} \;$Hz; \ (2) \ $V=0.998c\;$ and $\;b=6\times 10^{14} \;$Hz; \ (3) \
$V=0.9992c\;$ and $\;b=2.4\times 10^{14} \;$Hz].\ In all cases the aperture radius is
$R=4\;$mm. \ Remember the linear relation $\omega = V k_z + b$. \ The continuous lines are obtained from the closed-form
analytic expression (eq.23 in Ref.[12], while those represented by
dotted lines come from the numerical simulation of the
Rayleigh-Sommerfeld formula (eq.8 in [12]). The agreement is excellent: Practically, no
difference is perceivable.}
\label{figTNDP2.pdf}
\end{figure}

\begin{figure}[!h]
\begin{center}
 \scalebox{1.25}{\includegraphics{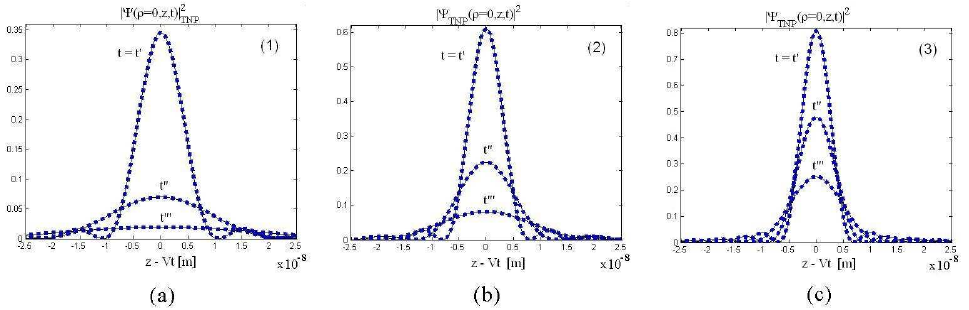}}
\end{center}
\caption{(Color online) On-axis evolution of the same subluminal TNDP considered in Fig.\ref{figTNDP2.pdf},
at three different instants of time [$t'=0.11\;$ns, $t''=0.22\;$ns and $t'''=0.33\;$ns]. \ Figures a, b and c represent the cases
(1),(2) and (3) respectively. \ The continuous lines are the
results obtained from the closed-form expression, while those represented by dotted lines come from
the numerical simulation of the Rayleigh-Sommerfeld formula. Again, the agreement is excellent: No difference is detectable.}
\label{figTNDP3.pdf}
\end{figure}

First of all, let us recall that in Ref.[12] it was developed a preliminary method for
describing the on-axis space-time evolution of truncated
non-diffracting pulses, be they subluminal, luminal or
superluminal. \ Within that quite simple method, the on-axis evolution of a TNDP depends only on the
frequency spectrum $S(\om)$ of the corresponding INDP $\Psi_{{\rm INDP}}$;
contrarily to the Rayleigh-Sommerfeld formula which
depends on the explicit mathematical expression of $\Psi_{{\rm INDP}}$. \ Such a heuristic method, due to its simplicity,
can yield closed-form expressions which describe the on-axis evolution of innumerable TNDPs.
In Ref.[12] one can find the analytic expressions for the truncated versions
of several well-known localized pulses: subluminal, luminal, or superluminal. Therein, the theoretical results were
compared with those obtained through the
numerical evaluations of Rayleigh-Sommerfeld integrals, with excellent agreement. Here, we
confine ourselves just to present an example of such noticeable agreements, by Figures \ref{figTNDP2.pdf} and
\ref{figTNDP3.pdf}.

\

\h The mentioned approach in [12] is actually useful, because in general it furnishes
closed-form analytic expressions avoiding the need
of time-consuming numerical simulations; and also because those closed-form
formulae provide an efficient tool for exploring several
properties of the truncated localized pulses: as their
depth of field, longitudinal pulse behavior, decaying
rates, etc.

\h However, let us turn to a more rigorous approach.

\subsection{Again on closed-forms for Non-Diffracting Pulses, in the Fresnel regime, generated by
finite apertures}

Let us fix our attention to the method developed in [9], that we shall call for brevity ``the MRB method"; it can
be found summarized, now, also in [13]. By that method, we got the analytic description of some monochromatic waves:
namely, of a few (important) beams generated by finite apertures.  The important point is that one can generalize the efficient method MRB, in the paraxial approximation, for the case of {\em pulses}.

\h Since we are going to use superpositions
of Bessel-Gauss beams, let us start by recalling the form of the so-called Bessel-Gauss beam[14]:

\bb \Psi_{\rm BG}(\rho,z) \ug -\frac{i k A}{2 z Q}\,e^{i k(z + \frac{\rho^2}{2z})} \, J_0\left(\frac{i k \kr \rho}{2zQ}\right)
e^{-\frac{1}{4Q}(\kr^2 + \frac{k^2\rho^2}{z^2})} \ ,
\label{bg} \ee

which appears to be a Bessel beam transversally modulated by the Gaussian function.  \ Quantity $Q=q - ik/2z$, and $\kr$
(the transverse wavenumber associated with the modulated Bessel beam) is a constant.  \ [When $q=0$, the Bessel-Gauss beam
results in the well-known Gaussian beam. \ The Gaussian beam, and Bessel-Gauss', Eq.(\ref{bg}), are among the few
solutions to the Fresnel diffraction integral that can be obtained analytically]. \ The situation gets
much more complicated, however, when facing beams truncated in space by finite circular apertures: For instance, a Gaussian
beam, or a Bessel beam, or a Bessel-Gauss beam, truncated via an aperture with radius $R$. [In this case, the upper limit of the Fresnell integral becomes the aperture
radius, and the analytic integration becomes very difficult, requiring recourse, as we were saying, to lengthy numerical calculations]. \ Afterward, let us also recall that --in the case of beams--  we considered the solution given by the following superposition of Bessel-Gauss beams

\bb \Psi(\rho,z) \ug - \frac{i k}{2 \, z}\,\,e^{i k(z + \frac{\rho^2}{2 \, z})} \dis{\sum_{n=-N}^{N}} \,
\frac{A_n}{Q_n}\, J_0\left(\frac{i \, k \,\kr \,\rho}{2 \, z \, Q_n}
\right)\,e^{-\frac{1}{4 Q_n}(\kr^2 + \frac{k^2 \rho^2}{z^2})} \; , \label{geral} \ee

quantities $A_n$ being constants, and $Q_n$ being given by \ $Q_n = q_n - \frac{ik}{2z} \label{Qn}$, \
where the $q_n$ are constants that can assume complex values.  In this superposition all
beams possessed the same value of $\kr$. \ In our previous work, we wanted the solution (\ref{geral}) to be able to represent beams truncated by circular apertures, in the case of Bessel beams, gaussian beams, Bessel-Gauss beams, and plane waves. \ And,
given one of such beams, truncated at $z=0$ by an aperture with radius $R$, we determined the coefficients
$A_n$ and $q_n$ in such a way that Eq.(\ref{geral}) represented with
fidelity the resulting beam.  More details in the papers of ours quoted above.

\h Let us recall, before going on, that in previous work we found an equation which
could be used for representing, on the plane $z = 0$, truncated Gaussian, Bessel, Bessel-Gauss beams
and truncated Plane waves; with the consequence that the evolution of such truncated beams was given
by Eq.(\ref{geral}). \ {\em The interesting question, for us, is now: Is it possible to derive from what precedes also analytic descriptions of} pulses truncated {\em by finite apertures?:} For instance, for TBP (truncated Bessel pulses), TBGP (truncated Bessel-Gauss pulses), TGP (truncated gaussian pulses), and TPP (truncated plane-wave pulses)?  [even if we shall fix our attention only on truncated Bessel pulses]. \ We shall answer this question within
the paraxial approximation; to this aim, consider an envelope $\psi(x,y,z;t)$ obeying the equation of the paraxial waves

\bb
\dis{
{{\pa \psi} \over {\pa z}} + {1 \over c} {{\pa \psi} \over {\pa t}} - {1 \over {2k}} {\nabla_\perp}^2 \psi \ug 0 }  \; ,
\label{eq02} \ee

where the time dependence of $\psi$ is essential [and cannot be eliminated as in the case of beams].
When assuming axial symmetry, one can write:

\bb
\psi(\rho,z,t) \ug \int_0^\infty d\kr \; \int_{-\infty}^\infty d u \, S(\kr,u) \, J_0(\kr \rho) \; e^{i {u \over c} z} \;
e^{-i {{\kr^2} \over {2k}} z} \; e^{-i u t} \; ,
\label{eq12} \ee

where we replaced $\omega$ with
the variable $u$ (since $\omega \equiv \omega_0$ will mean here the pulse central frequency). \ As usual, it is
\ $k \equiv  {\omega \over c}$, \ and \ $k_z \ug {u \over c} - {{\kr^2} \over {2k}}$. \ If we know $\psi(\rho,z,t)$ {\em on the plane} $z=0$ {\em of the aperture}, it will be

\bb
\psi(\rho,0,t) \ug \int_0^\infty d\kr \, \kr \; \int_{-\infty}^\infty d u \, S(\kr,u) \, J_0(\kr \rho) \;
 e^{-i u t} \; ,
\label{eq13} \ee

in terms of one Fourier-Bessel and one Fourier transformation. \ By using the corresponding inverse
transformations, one succeeds in writing the spectral function $S(\kr,u)$ as a function of the field existing
at the aperture!; namely:

\bb
S(\kr,u) \ug {1 \over {2 \pi}} \; \int_0^\infty d\rho \, \rho \; \int_{-\infty}^\infty d t \, J_0(\kr \rho) \;
 e^{i u t}  \; \psi(\rho,0,t) \; .
\label{eq14} \ee

Inserting Eq.(\ref{eq14}) into Eq.(\ref{eq12}), and reversing the order of the integrations over $k_\rho$ and $u$,
one gets

\bb
\begin{array}{clr}
& \psi(\rho,z,t) \ug  \int_0^\infty d\rho' \; \int_{-\infty}^\infty d t' \  \\

\\

& \times  \left\{ \ \rho' \psi(\rho',0,t') \left[ \int_0^\infty d\kr \kr \, J_0(\kr \rho) \, J_0(\kr \rho') \, e^{-i {{\kr^2} \over {2k}} z} \right]
\ \left[ \int_{-\infty}^\infty d u \, e^{i {u \over c} (z-ct)} e^{i u t'} \right] \right\} \ .
\end{array} \label{eq16} \ee

The last part in square brackets yields the Dirac delta \ $2 \pi \, \delta \left( t' + {{z-ct} \over c} \right)$. \
With some more algebra, one reaches the equation

\bb
\begin{array}{clr}
& \psi(\rho,z,t) \ug \int_0^\infty d\rho' \; \int_{-\infty}^\infty d t' \  \{ \rho' \psi(\rho',0,t') \\

\\

& \times  \left[ -i {k \over z} \; \exp \left[ i {{k \rho^2} \over {2z}} \right] \; \exp \left[ i {{k {\rho'}^2} \over {2z}}
\right] \; J_0 \left( {{k \rho \rho'} \over z} \right) \right] \ \left[ 2 \pi \; \delta \left( t' + {{z-ct} \over c} \right) \right] \ \} \ ,
\end{array} \label{eq19} \ee

which can be finally integrated over $t'$, without difficulties due to the presence of the delta, {\em furnishing for a pulse
the solution we were looking for:}

\bb
\begin{array}{clr}
\psi(\rho,z,t) & = \; \dis{ -i {k \over z} \; \exp \left[ i {{k \rho^2} \over {2z}} \right] } \\

\\

& \times  \dis{ \int_0^\infty d \rho' \rho' \; \psi \left( \rho',0,-{{z-ct} \over c} \right) \; \dis{ e^{i {{k {\rho'}^2} \over {2z}}} } \; J_0 \left( {{k \rho \rho'} \over z} \right) } \ ,
\end{array} \label{eq20} \ee

where one can notice that under the integral it now appears quantity $-(z-ct) / c$ instead of $t$.
\ Equation (\ref{eq20}) is the analogous of the one found out by our MRB method for beams.

\h { \em The integral solution (\ref{eq20}) tells us that the pulsed field (envelope) can be obtained by merely
knowing its value in the plane $z=0$ of the aperture, as a function of time and of the spatial coordinate.} \ The
result in Eq.(\ref{eq20}) is interesting also because it extends the MRB method to pulsed fields:
In the sense that one can utilize any solution found by the said method[9] for beams, transforming it
into a solution for pulses via a mere multiplication by the function
$\exp [ -(z-ct^2) / (c^2T^2) ]$. More precisely:\hfill\break

---to get a truncated beam, it is enough to have at the aperture a field of the type
\ $\psi(\rho,0) \approx J_0(\kr \,\rho) \ \sinc (\rho / R)$; \hfill\break

---to get a truncated pulse, it will be enough to have at $z=0$ an analogous field of the type
\ $\psi(\rho,0) \approx J_0(\kr \,\rho) \, \exp (-t^2/ T^2) \ \sinc (\rho / R)$; \hfill\break

where the multiplying function  \ $\exp [-(z-ct^2) / (c^2T^2)]$ \ reduced to  \ $\exp (-t^2/ T^2)$ \ on supposing that
$T >> 2 \pi / \omega$.  \ One can also notice that, having recourse to multiplying functions of the type \ $\exp [-(t/T)^{2n}]$, \
one can get a series of (for instance) step-shaped pulses.

\

\section{Applications for acoustic (ultrasonic)\hfill\break non-diffracting pulses}

Let us finally consider ultrasonic (acoustic) pulses, for instance with a central frequency
of 40 kHz, generated by a finite aperture with radius $R = 0.5$ m. \ One may have in mind, for example,
remote sensing, and the purpose of obtaining a realistic pulse which keeps its spot-size unvaried for, say, 20 m.
We shall apply of course the results of our last subsection, which allow us to describe analytically several
truncated pulses without any need, again, of lengthy numerical simulations.

\begin{figure}[!h]
\begin{center}
 \scalebox{4.8}{\includegraphics{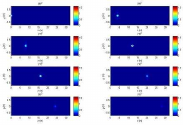}}
\end{center}
\caption{ (Color online) As an application of our last method to ultrasonic (acoustic) pulses, let us consider a truncated Bessel pulse with initial spot-radius of 15 cm, generated by a finite antenna with radius $R=0.5$ m.  It is here neglected its attenuation (actually
strong in the air when its central frequency is assumed to be 40 kHz).  If one has in mind, for example,
remote sensing, he may want our realistic pulse to keep its spot size unvaried for, say, 20 m. \ The present set of figures,
representing by colors the actual pulse evolution, does indeed show such a behavior.}
\label{BP1}
\end{figure}

\begin{figure}[!h]
\begin{center}
 \scalebox{5.17}{\includegraphics{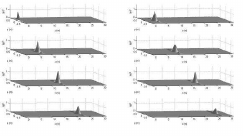}}
\end{center}
\caption{This second set of figures shows, in terms of 3D plots, the evolutions of the same truncated Bessel pulse considered in the previous Figure. \ In this case, the intensity is given by the height of $|\psi|^2$). \  From these figures
one can easily see the pulse spot (initially with a radius of 15 cm) to keep rather well its size for about 20 m, just with an
oscillating intensity due to the edge-effects of the finite antenna. \ Afterward, the pulse strongly deteriorates; and, to get better results by a Bessel pulse like this, one ought to use larger antennas.} \label{BP2}
\end{figure}

\begin{figure}[!h]
\begin{center}
 \scalebox{2.75}{\includegraphics{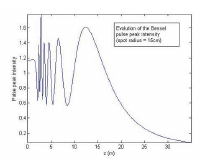}}
\end{center}
\caption{ (Color online)  In this further figure we depict the evolution, during propagation, of the intensity peak of the same truncated Bessel pulse; still keeping no account of attenuation.} \label{BP3}
\end{figure}

\h We shall confine ourselves, however, to just a few example. Let us start with a
truncated Bessel pulse with spot-radius of 15 cm. For simplicity, we shall not take here into account
the pulse attenuation, quite present for the said frequency in the air, even if one could take account of
it without too much difficulty.

\h Such a Bessel pulse, a priori, can be easily generated. If we think in terms of a simple antenna, constituted by
an array of annular transducers, then: (i) transducers do exist working with the mentioned
frequency; (ii) amplitudes and phases of the vibrations are given as functions of the chosen pulse; (iii) the
pulsed excitation (a modulation of the carrier wave) is the same for all transducers, and we choose precisely a temporal
gaussian with $\Delta t = 2.5 \;$ms, hundred times larger than the period of the 40 kHz wave ($T = 2.5 x 10^{-5} \;$ s).
Incidentally, the choice of pulses with duration much longer than the carrier period is requested by the
slow-envelope approximation, assumed by us when generalizing the MRB method for pulses.

\h Let us give an idea of the results by the help of suitable Figures.

\h The first set of figures, Fig.\ref{BP1}, shows the pulse evolution by colors. \ By contrast, the second set of figures,
Fig.\ref{BP2}, shows it in terms of 3D plots (the intensity being represented by te eigth of $|\psi|^2$). \  From figures \ref{BP2}
one can clearly see the pulse spot (initially with a radius of 15 cm) to keep rather well its size for about 20 m, just with an
oscillating intensity due to the edge-effects of the finite antenna. Afterward, the pulse strongly deteriorates; and, to get better results by such a Bessel pulse, one ought to use larger antennas.

\h By the the last figure, Fig.\ref{BP3}, we depict the evolution of its intensity peak while propagating (still keeping no account of the attenuation).

\

\

\subsection{Further Cases}

\

\

\h {\bf Second case: Truncated Bessel beam with a spot radius of 23 cm}

\

\noindent Suppose we want to get now a spot keeping its size for a larger distance, arriving at about thirty meters; while the ray
of the generating antenna remains $R=0.5$ m.

\begin{figure}[!h]
\begin{center}
 \scalebox{1.8}{\includegraphics{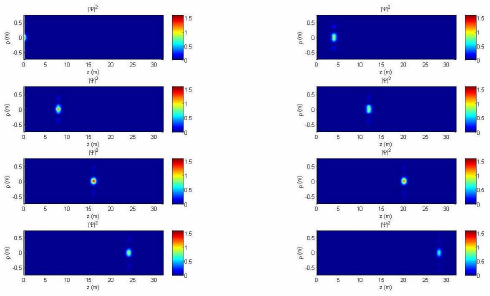}}
\end{center}
\caption{ (Color online) As a second application of our method to ultrasonic (acoustic) pulses, let us consider now a truncated Bessel pulse with an initial spot-radius of 23 cm, generated by a finite antenna still with radius $R=0.5$ m.  It is again neglected its attenuation (actually strong for a 40 kHz frequency).  We wanted our realistic pulse to keep its spot size unvaried for the larger distance of about 30 m. \ The present set of figures,
representing by colors the actual pulse evolution, does indeed show such a behavior.}
\label{BTII-1}
\end{figure}

\begin{figure}[!h]
\begin{center}
 \scalebox{1.9}{\includegraphics{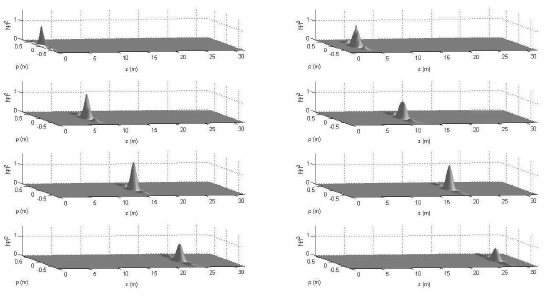}}
\end{center}
\caption{This second set of figures shows, in terms of 3D plots, the evolutions of the same truncated Bessel pulse considered in the previous Figure (its intensity being now the height of $|\psi|^2$). From the present figures one can see even better that the pulse spot (initially with a radius of 23 cm) keeps
rather well its size for almost 30 m.  \ Afterward, the pulse strongly deteriorates; and, to get better results by a Bessel pulse
like this, one has to use larger antennas.} \label{BTII-2}
\end{figure}

\h As before, the pulse evolutions is first shown by colors (Fig.\ref{BTII-1}),
and then in terms of 3D plots (Fig.\ref{BTII-2}), when the intensity is represented by the height of $|\psi|^2$.

\h One can see that the impulse spot radius (initially of 23 cm) maintains its value for about 30 m, oscillating in intensity
due to the edge effects of the finite antenna. \ Afterward, the pulse strongly deteriorates; once more, to get better results by a Bessel pulse like this, one ought to use larger antennas.

\begin{figure}[!h]
\begin{center}
 \scalebox{2.75}{\includegraphics{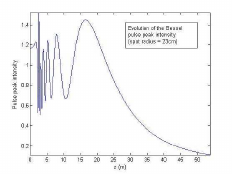}}
\end{center}
\caption{ (Color online)  In this further figure we depict the evolution, during propagation, of the intensity peak of the same truncated Bessel pulse.} \label{BTII-3}
\end{figure}

\h By the the last figure, Fig.\ref{BTII-3}, we depict the evolution of its intensity peak while propagating (when attenuation is neglected).

\

\

\newpage

\h {\bf Third case: Truncated Plane Wave Pulse}

\

\begin{figure}[!h]
\begin{center}
 \scalebox{1.98}{\includegraphics{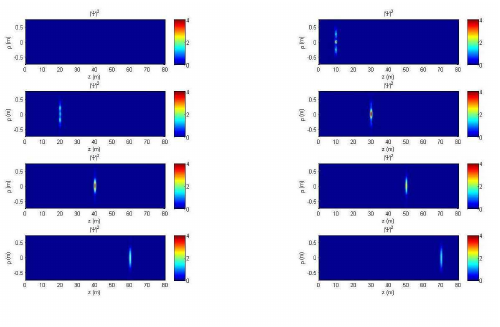}}
\end{center}
\caption{ (Color online) As a third application of our method to ultrasonic (acoustic) pulses, let us now consider a Truncated Plane Wave Pulse, with initial spot-radius of 0.5 m.  \ It is interesting that the spot size does {\em diminish} during propagation, till a distance of 30 m.  Only afterward, it start increasing, and the pulse opens. \ The present set of figures shown such a
behavior by colors.}
\label{PT-1}
\end{figure}

\begin{figure}[!h]
\begin{center}
 \scalebox{1.9}{\includegraphics{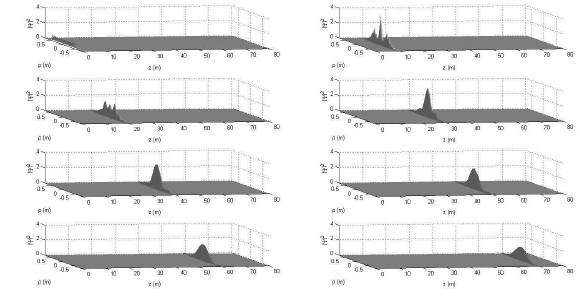}}
\end{center}
\caption{This second set of figures shows, in terms of 3D plots, the evolutions of the Truncated Plane Wave Pulse considered in the previous Figure. From the present figures one can see even better that the pulse spot (initially with a radius of 0.5 m) {\em reduces} its size for almost 30 m, reaching the value of 0.3 m.  \ Only afterward the pulse opens (its spot-size increasing).
} \label{PT-2}
\end{figure}

\noindent The first set of figures, Fig.\ref{PT-1}, shows the pulse evolution by colors; the second set, Fig.\ref{PT-2}, shows it by 3D plots (the pulse intensity being the height of $|psi|^2$).

\h In this case one clearly recognizes the interesting fact that the initial spot-radius, of 0.5 m, changes during propagation {\em diminishing} during the first 30 meters till 0.3 m. \ After such a distance, however, the pulse starts to open: and its spot-size
increases.

\begin{figure}[!h]
\begin{center}
 \scalebox{2.75}{\includegraphics{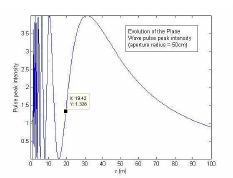}}
\end{center}
\caption{ (Color online)  In this further figure we depict the evolution, during propagation, of the intensity peak of the same Plane Wave truncated pulse.} \label{PT-3}
\end{figure}

\h By the the last figure, Fig.\ref{PT-3}, we depict the evolution of its intensity peak while propagating (when attenuation is neglected).

\

\

\

\

\

\

\h {\bf Fourth case: Pulse of Plane Wave Truncated and Focalized (at $z=20$ m)}

\

\begin{figure}[!h]
\begin{center}
 \scalebox{2.0}{\includegraphics{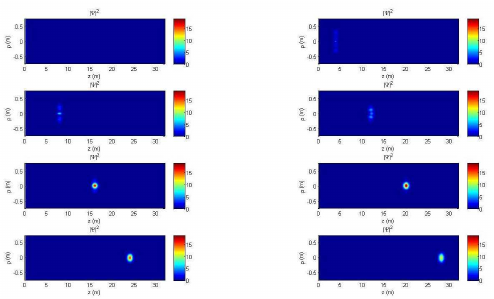}}
\end{center}
\caption{ (Color online) As a third application of our method to ultrasonic (acoustic) pulses, let us now consider a Focalized Plane Wave (Truncated) Pulse, with initial spot-radius of 0.5 m. \ In this case a focalization takes place at $z=20$ m.  Such quite interesting result has been obtained by having recourse to a Plane Wave truncated (and pulsed) at the aperture, and modulated by a phase function similar to the transfer function of a convergent lens with a 20 m focal distance. \ The present set of figures shown such a behavior by colors.}
\label{PF-1}
\end{figure}

\begin{figure}[!h]
\begin{center}
 \scalebox{1.89}{\includegraphics{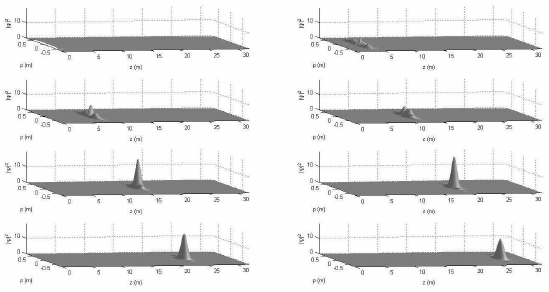}}
\end{center}
\caption{This second set of figures, in terms of 3D plots, shows the evolutions of the Focalized Plane Wave (Truncated) Pulse considered in the previous Figure. One can easily see that a focalization takes place at $z=20$ m. \ This interesting result has been obtained ---as we said--
by having recourse to a Plane Wave truncated (and pulsed) at the aperture, and modulated by a phase function similar to the transfer function of a convergent lens with a 20 m focal distance. \ The pulse leaves the antenna with a spot of 50 cm, which
shrinks down (while the intensity increases), till reaching the distance of 20 m where the spot gets its minimum radius, of
about 20 cm, and an intensity 20 times larger than the one at the aperture.
} \label{PF-2}
\end{figure}

\noindent The first set of figures, Fig.\ref{PF-1}, shows the pulse evolution by colors; the second set, Fig.\ref{PF-2}, shows it by 3D plots (the pulse intensity being the height of $|psi|^2$).

\h In this case one can clearly see that a focalization takes place at $z=20$ m. \ This quite interesting result has been obtained
by having recourse to a Plane Wave truncated (and pulsed) at the aperture, and {\em modulated by a phase function similar to the transfer function of a convergent lens with a} 20 m {\em focal distance}. The pulse leaves the antenna with a spot of 50 cm, which
shrinks down (while the intensity increases), till reaching the distance of 20 m where the spot gets its minimum radius, of
about 20 cm, and an intensity 20 times larger than the one at the aperture.

\begin{figure}[!h]
\begin{center}
 \scalebox{2.75}{\includegraphics{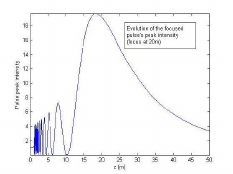}}
\end{center}
\caption{ (Color online)  In this further figure we depict the evolution, during propagation, of the intensity peak of the same Focalized Plane Wave (Truncated) Pulse.} \label{PF-3}
\end{figure}

\h By the the last figure, Fig.\ref{PF-3}, we depict the evolution of its intensity peak while propagating (when attenuation is neglected).

\

\

In the coming Section we are going to propose acoustic (ultrasound) antennas suitable for detection (or explosion)
of buried objects, like mines.

\

\newpage

\newpage

\
\begin{center}


\section*{Supplementary material}

# 3 – Towards the Proposal of Acoustic Antennas for Detection (or Explosion) of under-ground buried objects, like Mines.

One of our main aims in this work is the proposal of antennas, producing Acoustic Localized Waves (better called ANDW = Acoustic Non-Diffracting Waves), for the detection of buried objects, like MINES, or for causing their explosion from a distance of at least 10 meters.

Let us recall, before all, that the Localized Waves are non-diffracting, and propagate in a single direction without deformation, with energy concentrated within a spot. Such good properties are kept for infinite times and lenghts, as we know, only by ideal beams; in the case of realistic beams, produced by finite antennas, those properties are maintained for a certain (finite, even if long) depth of field. Let us recall also the following: (i) the majority of the ANDWs considered in the literature are the "X-shaped" supersonic ones, initially produced in 1992 [6] in analogy to what had been theoretically predicted[15], and soon concretely utilized for an ultrasound scanner which directly furnishes 3D, high-resolution images of the human body[16], in particular of moving organs like the heart. (In the electromagnetic case, the X-shaped waves resulted to be endowed with superluminal peak-velocities, in hundreds of theoretical, mathematical, numerical simulation, and experimental works[1,2] ); (ii) Subsequently, they have been investigated also Non-Diffracting Waves [NDW] with subluminal, in the electromagnetic case, and subsonic, in the acoustic case, peak-velocities; (iii) Particularly interesting they resulted to be the subluminal NDWs with zero peak-velocity: that is, with a static envelope (within which only the carrier wave propagates). They have been called *Frozen Waves* [FW], corresponding to electromagnetic, or acoustic, fields "at rest"[17]. The FWs can be created within the desired (even quite small) space region, with the chosen shape and intensity[18,19]. They are generated by superposing Bessel beams with the same frequency, which therefore can be prefixed too.


## 3.1 – Proposal of antennas creating *Acoustic Frozen Waves* [AFW]

Let us propose now the generation of *Acoustic Frozen Waves*, without forgetting that the air is an absorbing medium, especially for ultrasound waves. In the case of FWs we constructed a new theoretical *method* for their production, which allows to create the desided "static" field even in a lossy medium; in other words, it [20,21] allows *producing beams resisting to both the effects of diffraction and attenuation.* For the reader's convenience, it will be summarized in the Appendix

Let us start by depicting how a single Bessel beam [Bb] behaves in an *absorbing* medium. For instance, let us consider an acoustic Bb of 100 kHz, with an initial spot of 1.5 cm. For such

a frequency, the Bb is subject in air to an attenuation coefficient of 2 dB per meter, and will therefore possess a field depth of 2 meters only. The two following figures, Figs.15, show such an effect for an isolated Bessel beam; the square magnitude $|\Psi|^2$ of its field intensity is shown in figure (a), as a function, besides of the transverse co-ordinate $\rho$, of the co-ordinate z along the propagation axis: Both co-ordinates being expressed in meters. Figure (b) depicts its ortogonal projection

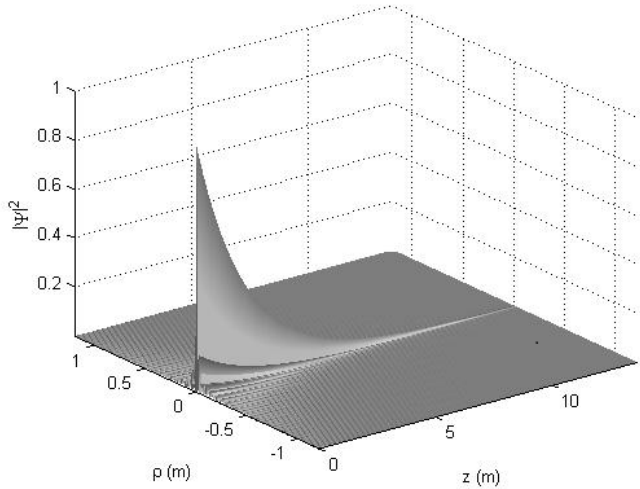

(15 a)

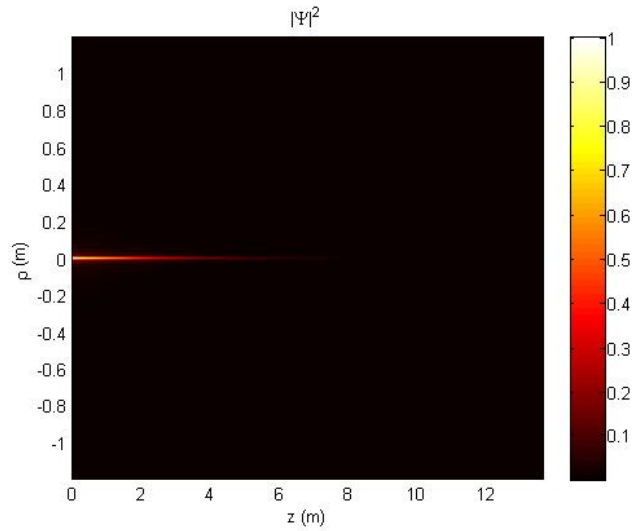

(15 b)

Figs.15: The case of an isolated acoustic Bessel beam (100 kHz; spot of 1.5 cm), subject in air to strong attenuation (2 dB/m), so that its field depth is expected to be 2 m.   Figure a):   3D behaviour of the field-intensity square-magnitude  $|\Psi|^2$ as a function of ($\rho$ and) z.   Figure b): orthogonal projection.

## First Example:

Let us choose the frequency of 100 kHz, in which case the attenuation in air is 2 dB/m. For a normal beam (plane wave, or gaussian beam) the depth of field would be of 2.2 m.  We know that by suitable superposition of equal-frequency Bessel beams[20,2] we can obtain FWs resisting to attenuation (besides to diffraction).

   The beam we want to obtain is a FW with a field depth of  11 meters, the acoustic field being moreover concentrated, at that distance, so to build a spot of 3 cm.  Our methodology[20,2] than suggests to be convenient an aperture of  0.9 m.  Let us require the intensity longitudinal pattern of the FW to be of a constant value over a rectangular region: we shall say it to be of a  "double step" type[20] in the interval  0< z <11 m.  More precisely, $\Psi(z)=1$  for  0< z <11 m, and  zero elsewhere.  When having recourse to a superposition of a small number of Bb's, the results are of the type presented in Figs.16

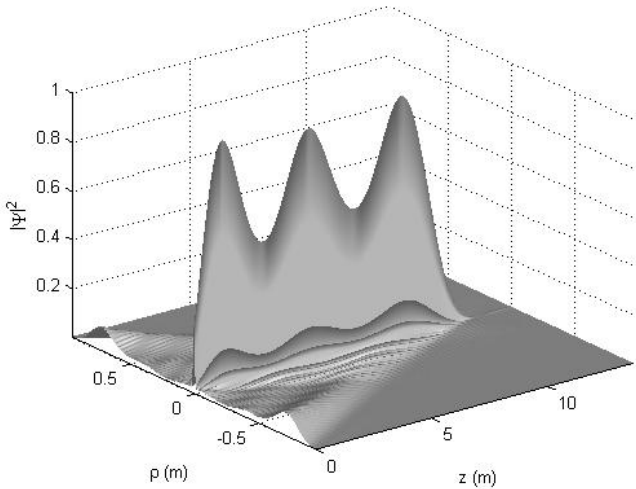
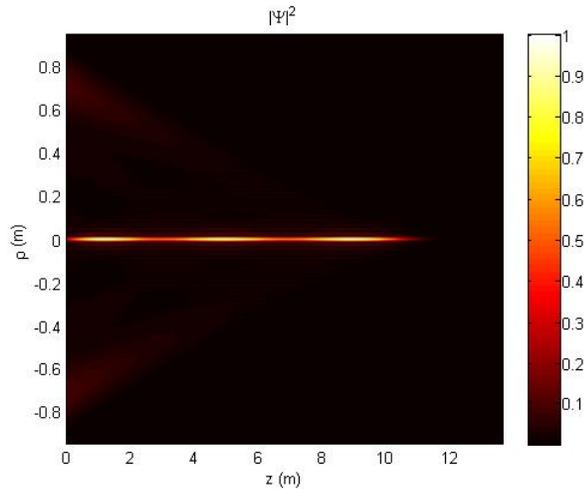

(16 a)    (16 b)

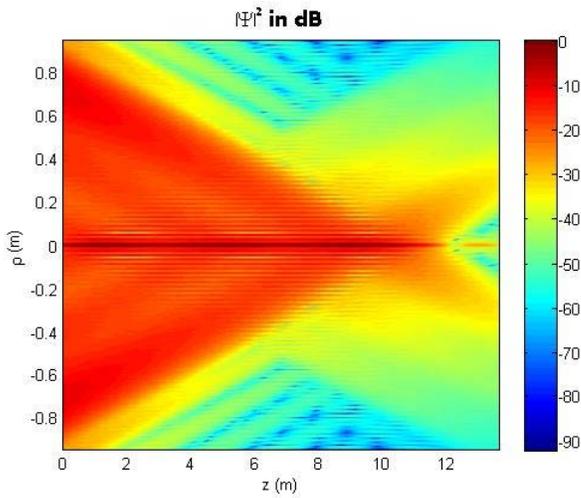
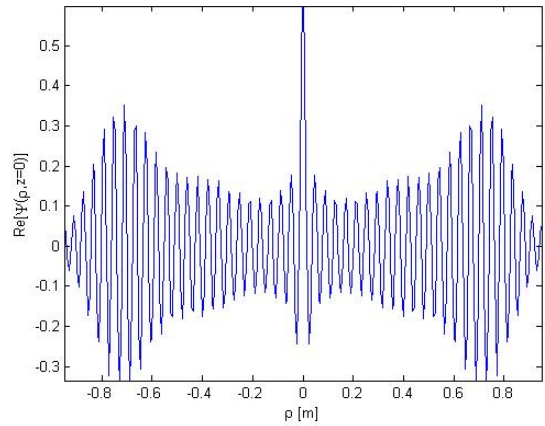

(16 c)  (16 d)

Figs.16: (a) Intensity behaviour of the FW in the case of the First Example (a "double step" shape, with a small number of Bessel beams in the superposition); (b) Orthogonal projection; (c) Orthogonal projection by a logarithmic scale plot (i.e., in dB); d) Real part of the field at the aperture (z=0). As requested, the field depth results to be 11 m, with a spot radius of 0.3 m.

## *Second Example:*

Let us consider a case analogous to the previous one (the longitudinal intensity shape being again of the "double step" type, in the interval $0< z <11$ m), but aiming this time at obtaining the maximum precision allowed by the method, which implies in the present case a maximum number of 101 Bb's in the superposition[20]. Still, the frequency be 100 kHz, and the attenuation coefficient in air of 2 dB/m. Let us recall the depth of field of a normal beam (plane wave or gaussian beam) to be of 2.2 m.

Our aim, once more, is getting a field depth of 11 m, with a spot diameter at that distance of 3 cm. We need an aperture of 0.9 m.

The results, by the new superposition, are the ones in Figs.17.

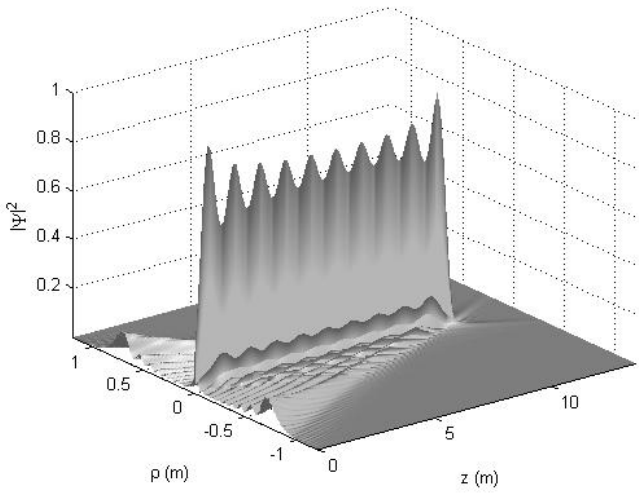

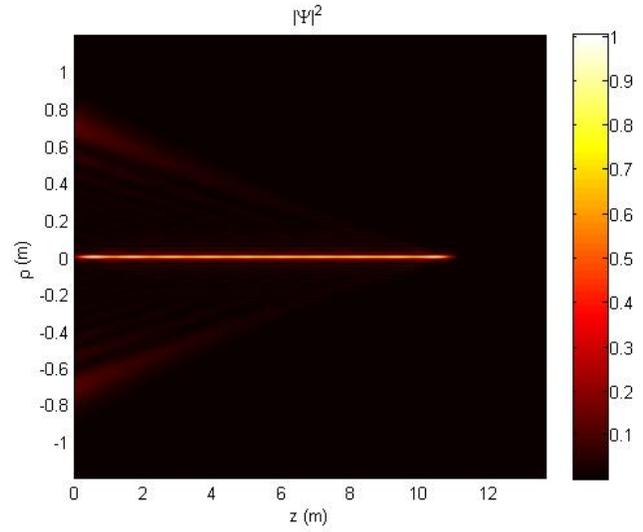

(17 a)

(17 b)

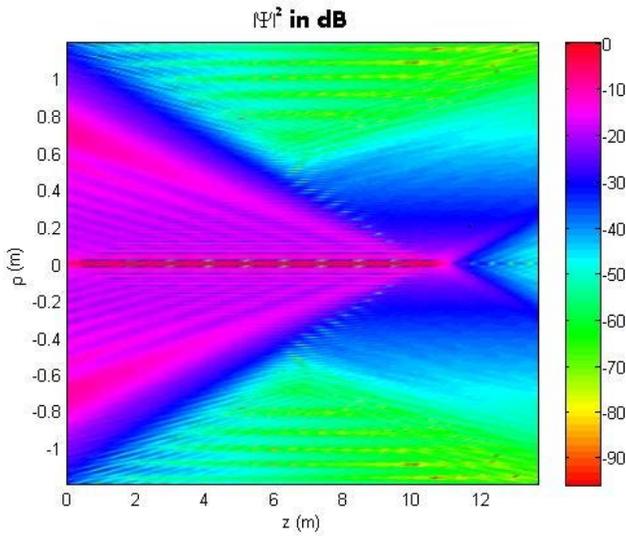

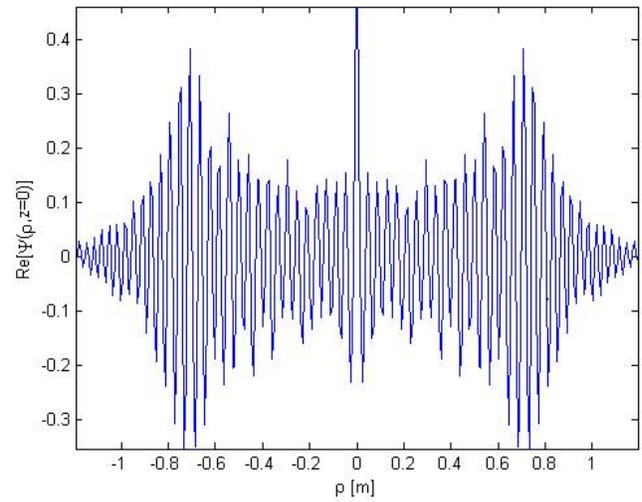

(17 c)

(17 d)

Figs.17: (a) Intensity behaviour of the FW in the case of the Second Example (still a "double step" shape, but with the maximum allowed number of Bb's in the superposition); (b) Orthogonal projection; (c) Orthogonal projection by a logarithmic scale plot (i.e., in dB); d) Real part of the field at the aperture ($z=0$). As requested, the field depth results to be 11 m, with a spot radius of 0.3 m, but this time with the maximum precision allowed by the method.

## *Third Example:*

Imagine we wish now to get a FW with a depth of field of 20 meters. The needed aperture radius becomes now of 1.4 m. The longitudinal intensity shape be again a "double step", but in the interval 0< z <20 m. Obtaining a field depth of 20 m brings us close to the limits of validity of our method, but we can still reach good results. The intensity of the lateral converging beams will increase, in order to reconstruct the resulting beam till a larger distance (20 m). The results appear in Figs.18.

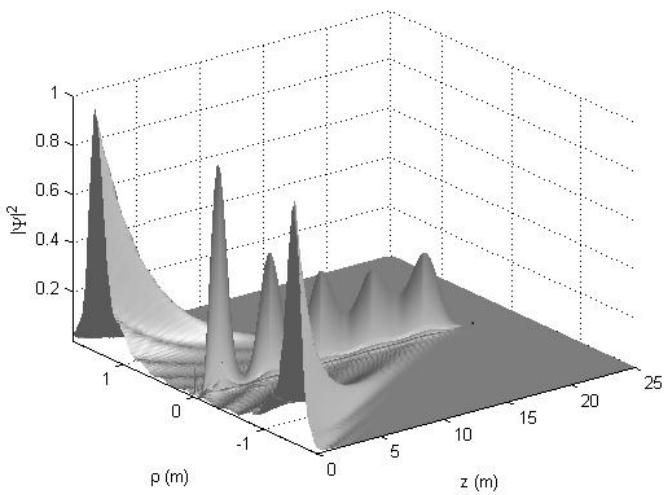
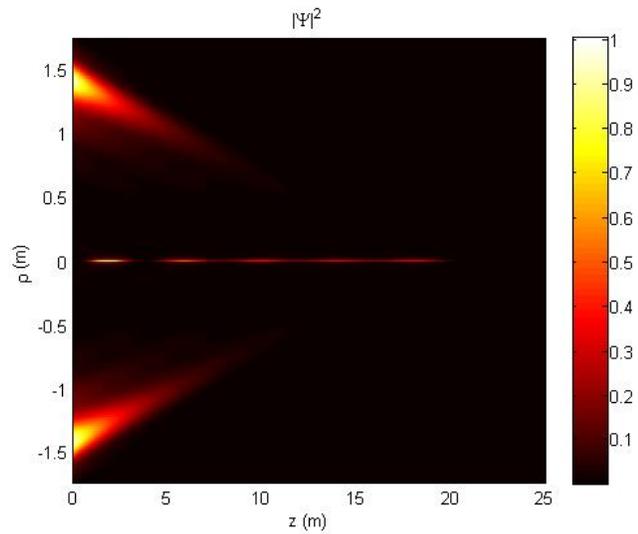

(18 a)                                           (18 b)

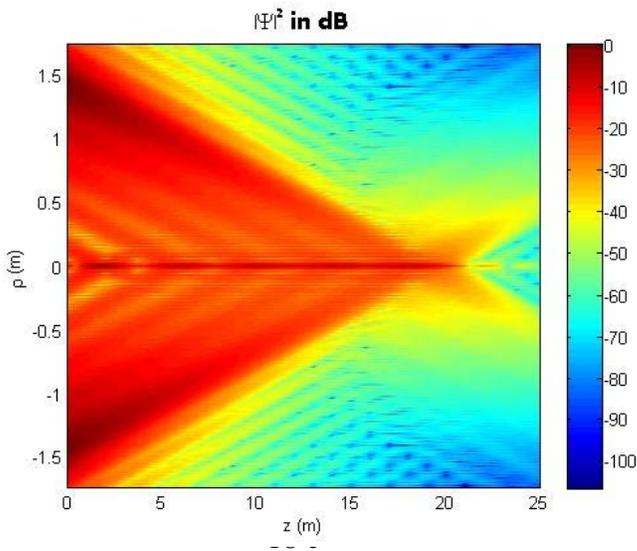
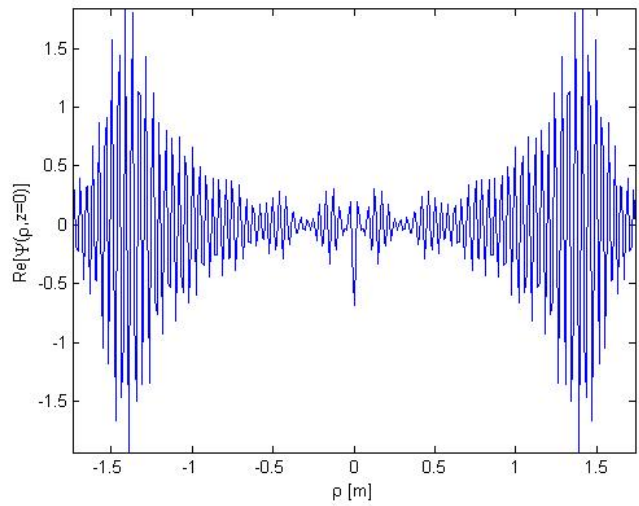

(18 c)   (18 d)

Figs.18: These Figures illustrate the 3rd Example, referring this time to a FW with a depth of field of 20 meters. The needed aperture radius then becomes of 1.4 m. The longitudinal intensity shape is again a "double step", but in the interval $0 < z < 20$ m. Requestin a field depth of 20 m brings us close to the limits of validity of our method, but we still reach good results. Figure (a): Intensity behaviour of the FW in the case of the Third Example; (b): Orthogonal projection; (c) Orthogonal projection by a logarithmic scale plot (i.e., in dB); d) Real part of the field at the aperture (z=0). As requested, the field depth results to be 20 m.

More material (equipped with suitable further Figures) will appear elsewhere.

# 4 – Some Considerations about ultrasonic *High-Power* Transducers for generating Non-Diffracting Beams or Pulses.

In this Section we are going to pay some preliminary attention to the construction of antennas for the generation of nao-diffracting (or even ordinary) *high-power* ultrasound beams, or pulses.

More specifically, we shall fix our attention --as before— to ultrasound beams resisting the effects of both diffraction and attenuation.

## 4.1 Introductory remarks

Before going on, let us however *recall* that the use of non-diffracting ultrasound beams, or pulses, is already a reality in the realm of medical imaging. Well known are the applications of acoustic X-shapes waves for high resolution ultrasound scanning, by Jian-yu Lu. Our own research group has been proposing the adoption of the methods of the ultrasound Frozen Waves for modeling non-diffracting beams in lossy media: in order to
approach diverse applications, from imaging to remote sensing, and to curing tumours by killing the cancer cells via heating produced by extremely localized ultrasonic beams.

For instance in [1] and [2], we suggested arrays of annular concentric transducers to create very-high resolution ultrasonic FWs (cf. Figs.19-21). Figure 19 is an example of acoustic annular aperture, with radius 24.5 mm, and 41 rings ($d$ = 0.5 mm; $\Delta_d$ = 0.1 mm): namely, it shows the section of an annular transducer array for constructing 1.5 MHz frozen waves. By such an array, we shall attempt producing, in amplitude and phase, the field of the chosen FW at the antenna location, that is, in the plane z = 0.

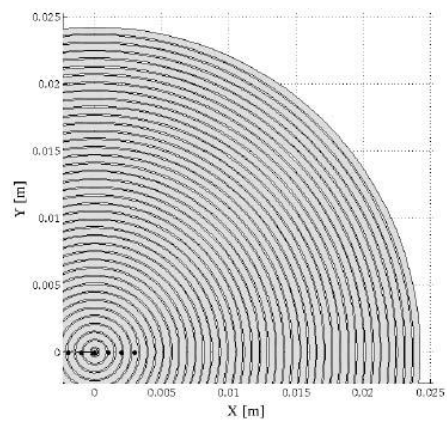

Fig. 19 – Example of acoustic annular aperture having a radius of 24.5 mm and 41 rings (with width $\Delta_d = 0.1$ mm, and inter-ring distance $d = 0.5$ mm). The figure shows the section of the annular transducer array chosen for producing FWs of 1.5 MHz.

Before that, let us show (by Fig.20 and Fig.21) a sample of the exitation pattern (amplitude and phase, respectively) corresponding to the array in Fig.19: With $R = 24.5$ mm, $N_r = 41$, and $f = 1.5$ MHz. On the horizontal axis it appears the radius (in meters) of the emitting ring (transducer).

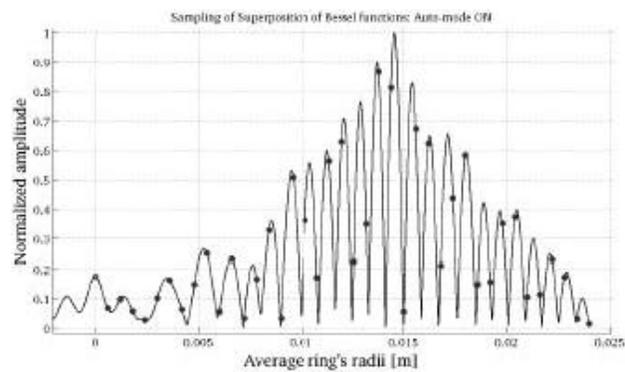

Fig. 20 – *Amplitude* pattern excited by the acoustic array in the previous Figure (radius of the array $R = 24.5$ mm; with 41 emitting rings, and ultrasonic frequency f = 1.5 MHz). The values, in meters, on the horizontal axis represent the radii of the various emitting rings (transducers).

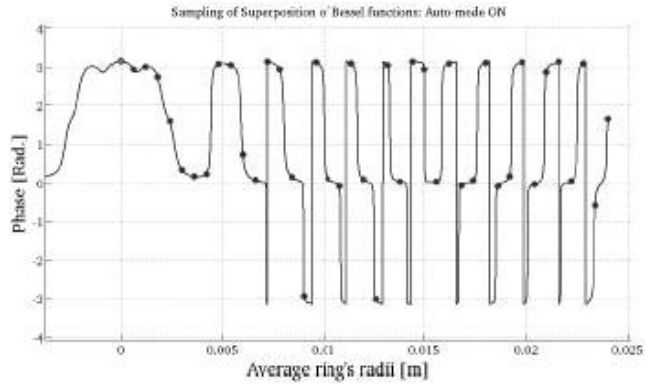

Fig. 21 – Pattern of the acoustic wave *phase* excited by the array in the previous Figure (that is, radius of the array *R* = 24.5 mm; with 41emitting rings, and ultrasonic frequency f = 1.5 MHz). The values, in meters, on the horizontal axis still represent the radii of the various emitting rings.

One should discretize the continuous field distribution at the aperture, as suggested and performed by us in [9], and in ref.[13], downloadable e.g. from http://arxiv.org/abs/1408.3635 , so that the annular transducers locations can be determined by such a discretization (see figures 7 and 12 therein).

Anyway, a possible set-up for the generation of FWs, when having recourse to an array of annular transducers, may be the one sketched in Fig.22.

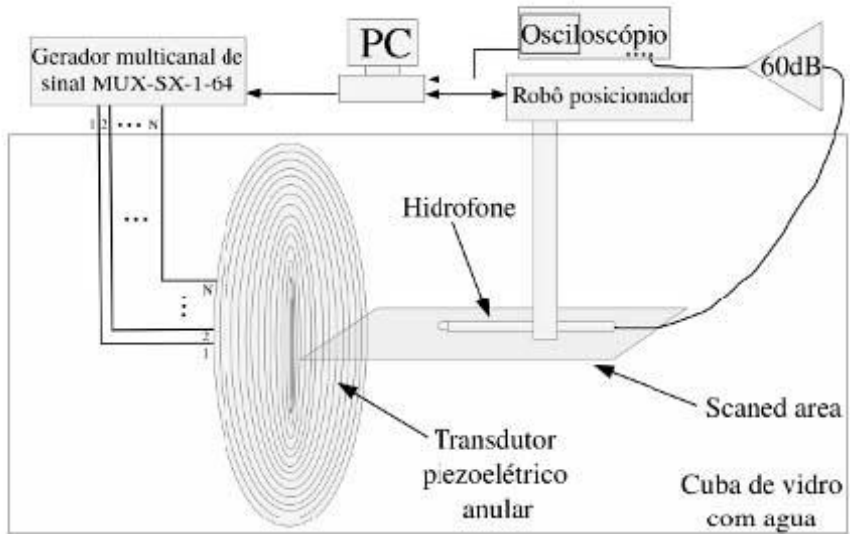

Fig. 22 – A possible experimental set-up for the creation of FWs, when having recourse to an array of annular transducers.

As a mere example, let us first forget about attenuation, and examine how a theoretical FW can be generated by use of an annular transducer antenna. One obtains the FW corresponding to the adopted array, by numerical simulation of the field emanated by it[3]. In the following Figures the values of z are in meters, while those of $\rho$ are in mm.

Let us operate with $f = 2.5$ MHz, and $L = 30$ mm. The theoretical FW, to be experimentally constructed, is chosen to consist in two rectangular regions with different intensities; and to be expressable as the superposition of $2N+1 = 25$ Bessel beams. One thus obtains an highly concentrated ultrasonic beam, with a spot smaller that 1 mm (near to the diffraction limit): as it is shown in our theoretical Fig.23.

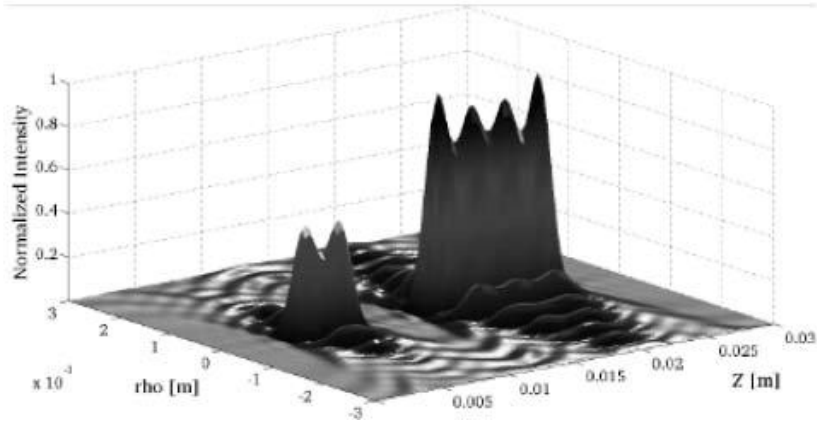

Fig. 23 – The theoretical FW we choose, for subsequent experimental creation, consists in two rectangular regions with different intensities; and is expressed as a superposition of $2N+1 = 25$ Bessel beams (we operate with $f = 2.5$ MHz, and $L = 30$ mm, neglecting attenuation: Cf. [3]). As depicted in this figure, one thus obtains a highly concentrated ultrasonic beam, with a spot smaller that 1 mm (near to the diffraction limit).

Figure 24 shows the result of the corresponding experiment, or rather in our case *simulated experiment,* for the generation of the chosen supersonic FW in Fig.23 (still disregarding attenuation). The annular aperture (with radius $R = 35$ mm) is located on the left side of Fig.24, at $z = 0$. The parameters of the simulation are $N_r = 101$ rings, again with $f = 2.5$ MHz, and $L = 30$ mm. The simulated experiment results to be rather satisfactory. All this

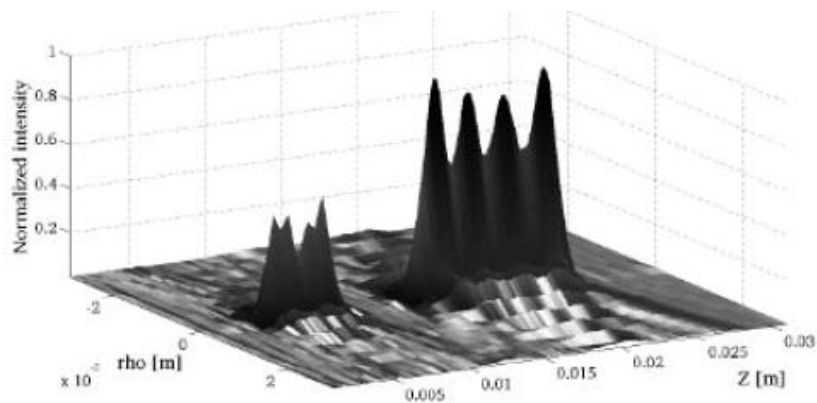

Fig. 24 -- Result of the supersonic *simulated experiment* aimed at constructing the FW appearing above in Fig.23. The annular aperture (with radius $R = 35$ mm) is located on the left side, at $z = 0$. Axis $\rho$ represents the transverse cylindrical co-ordinate, while z is the longitudinal one. The parameters of the simulation are: $N_r = 101$ rings, again with $f = 2.5$ MHz, and $L = 30$ mm. The simulated experiment results to be rather satisfactory.

does indicate that the experimental generation of non-diffracting beams, and more specifically, of FWs is a real possibility also in the ultrasonic region.

## 4.2  High-power ultrasonic non-diffracting beams or pulses

Let us eventually come to the main issue of this Section 4, which regards the creation

of non-diffracting ultrasonic beams, or pulses, having in mind particularly the detection from a distance of terrestrial mines, or even their explosion.

Two problems are to be faced:

(i) the strong attenuation in air of the ultrasonic waves. We saw in the previous Sections that, e.g, for the 100 kHz frequency, the attenuation is 2 dB/m, implying a quite short depth of field (2 m). We saw it even for a Bessel beam in the air;

(ii) the large power required for such purposes. Indeed, it is relatively easy to work out

 transducers for medical applications (imaging, etc.), but it is a  considerable

 challenge to generate very high power ultrasound.

Happily enough we can always obtain ultrasonic beams, resisting for long distances to both diffraction and attenuation, *by generalizing the method of ours for FWs* [20,4,2] (our standard method for FWs is briefly summarized in an Appendix, for possible convenience of the reader). Let us recall all that by exploiting here one more example of the type examined in Section 2 above.

### *One more Example (the Fourth)*

Similarly to what it was done in Section 2, let us now choose:

Frequency: 100 kHz

Attenuation coefficient in air: 2 dB/m

Depth of field of a normal beam (plane wave or gaussian beam) = 2.2 m

Requested field depth for the FW beam: 10 m

Radius of the desired spot: 5 cm

Needed aperture radius: 0.35 m

Longitudinal intensity pattern chosen: Of the "two steps" type in $0<z<10m$

Our theoretical method[20,4,2] allow us to costruct an acoustic Frozen Wave with such characteristics, as Figs.25 do show.

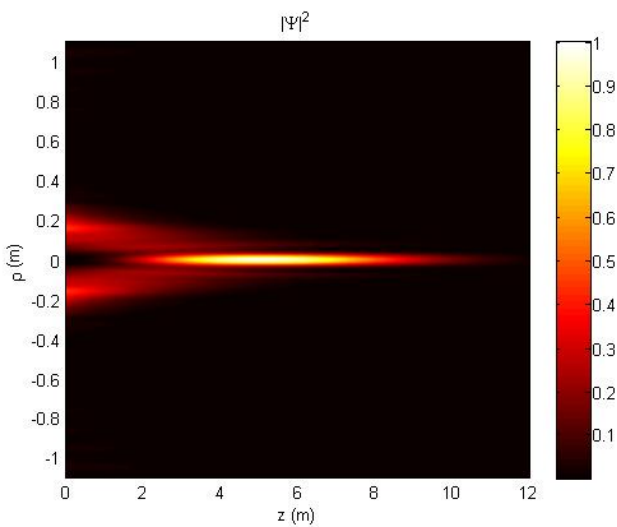
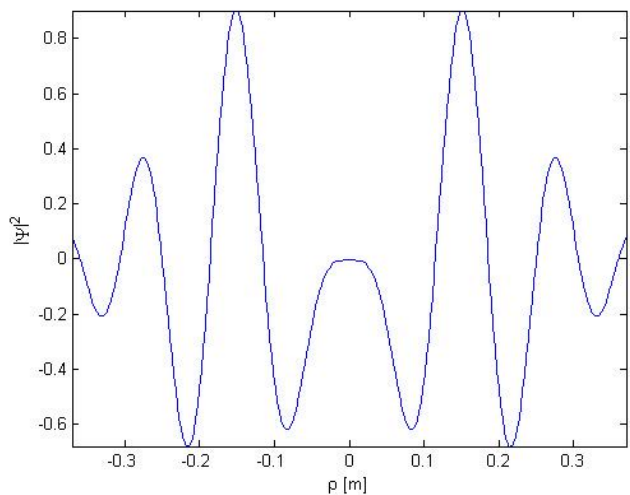

(25 a)　　　　　　　　　　　　　　　　　　　　　(25 b)

Fig. 25 – Our theoretical method[20,4,2] allows constructing an acoustic Frozen Waves with the characteristics specified in the *Example* above. This is shown in the present Figure.

It is rather encouraging the theoretical possibility of obtaining ultrasonic beams, resisting the effects of diffraction and attenuation for distances very much longer than the ones attained by ordinary beams[4]. But, let us repeat, the high-power needed for the desired applications, is often a serious obtacle.

In the case of the (last) Example above, one needs an array of annular transducers much bigger than the ones for medical applications.... In Fig.26 we show, e.g., a possible discretization: It is rather simple, but it it each point represents an annular transducer with at least 1 cm of thickness.

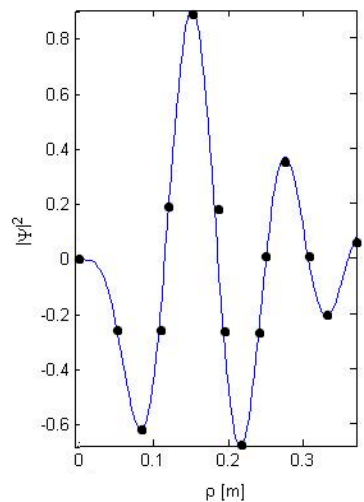

Fig. 26 – For the (last) example above, represented by Figs.25, we can suggest the discretization[9,13] appearing in this Figure: A very simple discretization, but in it each point indicates an annular transducer with at least 1 cm of thickness.

# 5 - Mentioning some available technology
   for the generation of high-power ultrasound

Let us suggest here, among the existing technologies, one which seems to be well developed, and is presented in a series of papers: We base ourselves, e.g., on the work[22] by Gallego-Juarez et al., appeared in 2010 in *Ultrasonics Sonochemistry* **17**, pp.953-964.

   One may then show how ultrasonic transducers of high-power may be constructed with large radiation areas, and with a design reducing non-linear effects and unwanted modal interactions.

   A basic structure for such transducers is the one in Fig.27 (a figure taken with permission from reference [22]): It consists essentially in a piezoelectric vibrator which drives a radiator. The vibrator is a sandwich of piezoelectric elements.

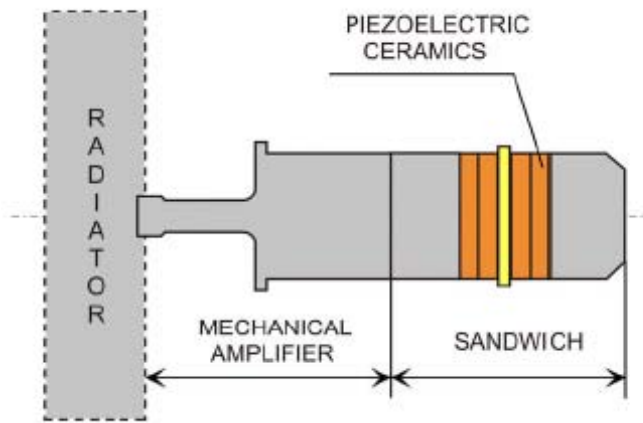

   Fig. 27 – Ultrasonic transducers of high-power may be constructed with large radiation areas and with a design that reduces non-linear effects and unwanted modal interactions.  A basic structure for such transducers is the one in the Figure (a figure taken with permission from reference [22]): It consists essentially in a piezoelectric vibrator which drives a radiator; the vibrator being a sandwich of piezoelectric elements.

The vibration caused by the piezoelectric elements is amplified by a mechanical amplifier, causing the radiator to vibrate in one of its modes. Such a radiator may have different geometries, varying with the application; in general, it can be circular or rectangular. Its surface can be moreover modeled, by furrows or bulges, so to get increased control on the radiated field (evoiding mode couplings, etc.) superfície pode possuir elevações e/ou sulcos, de tal forma a obter um certo controle sobre o padrão de radiação (avoiding mode couplings, etc..). The interested reader can be addressed, e.g., to figures 2-4, 9, 12-13, and 15 in ref.[22] for examples of circular transducers operaing in the 10--40 kHz region.

Let us here show the quite simple cases of Fig.28, reproduced with permission from [22].

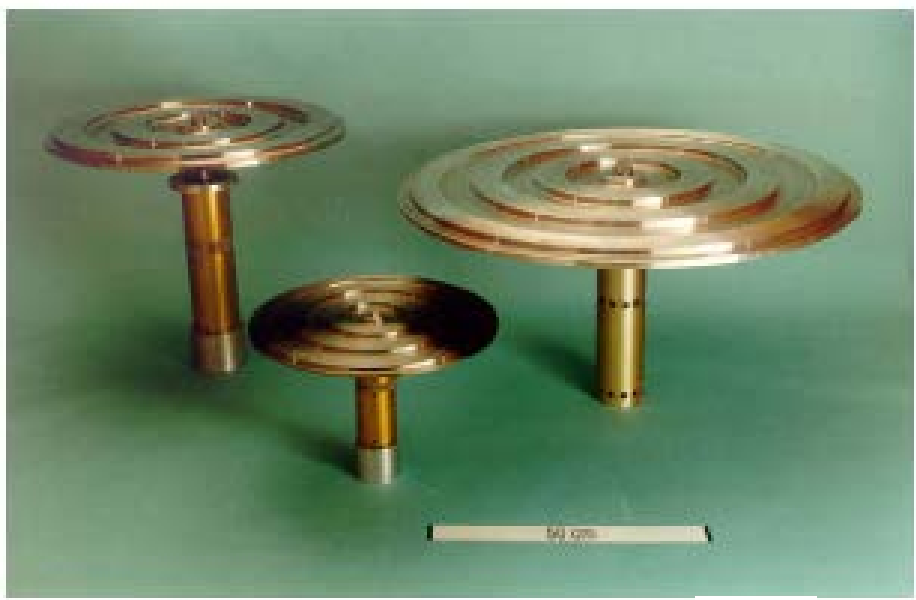

Fig. 28 – Simple circular radiators like those in this Figure (taken with permission from ref.[22]) are useful for creating common ultrasonic beams as the gaussian one; and may be sufficient also for a Bessel beam, under certain conditions. For more complex beams, like the FWs, more sophisticated arrays of high-power transducers are known to be needed.

Simple circular radiators like those in Fig.28 are useful for creating common ultrasonic beams as the gaussian one; and may be sufficient also for a Bessel beam (when higher order modes are excited, which implies, however, also higher frequencies). For more complex beams, like the FWs, more sophisticated arrays of high-power transducers are known to be needed. We are going back to such issues, having in mind the high-power generation of ultrasound beams (resisting diffraction and attenuation).

## 5.2 Possible arrays of high-power radiators for ultrasonic beams (resisting diffraction and attenuation)

We have seen that an efficient way for generating non-diffracting beams is the use of an array of annular transducers.

Now, we are passing to suggest for instance the use of *sets* of annular radiator arrays, by respecting (see below) annular symmetry; each constituent array being of the type in Fig.28.

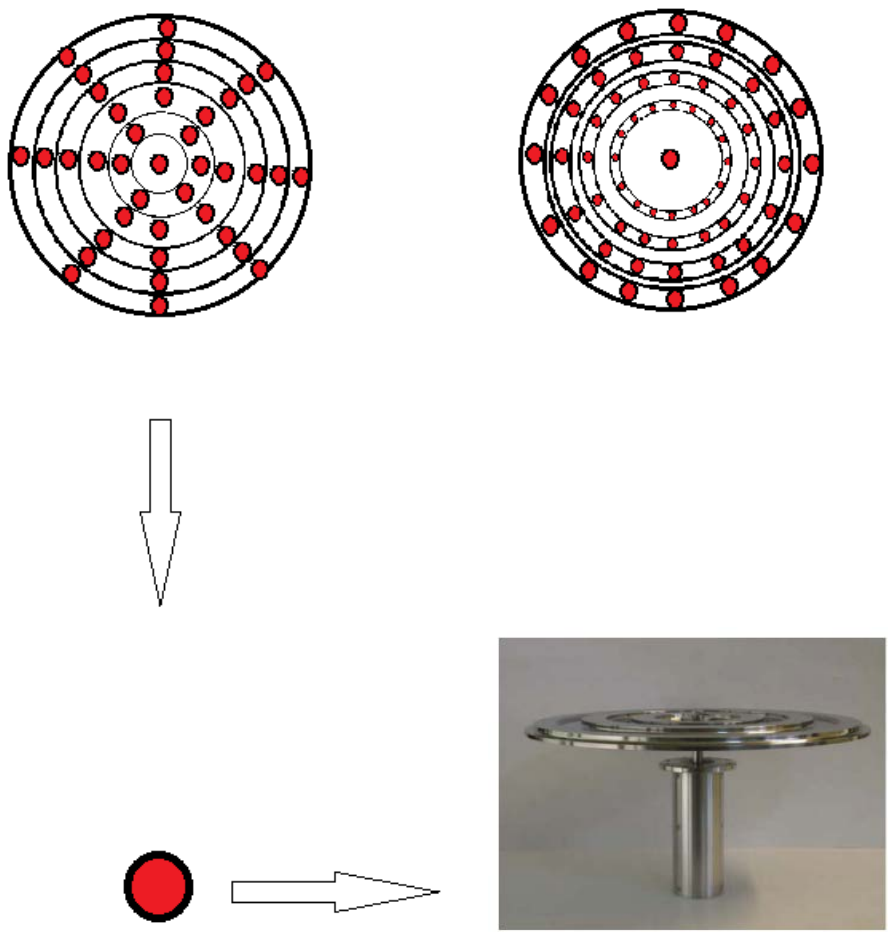

**Fig. 29** – For the generation of more sophisticated non-diffracting beams, let us now suggest for instance the use of *sets* of annular radiator arrays, by respecting annular symmetry: See the Figure, where it is clearly indicated that each constituent array is of the type

in Fig.28. The global radius of the set of radiators may be, e.g., of about 60—80 cm, while each radiator may have a diameter of 10—15 cm.

In Fig.29 the radius of the set of radiators may be, e.g., of about 60—80 cm, while each radiator may have a diameter of 10—15 cm.

A further possibility is discretizing the annular radiators via the introduction of rectangular (or bended) radiator elements. *Notice* that the exitation to be used for each radiating element can be obtained from the exact analytic solution which describes, for example, the chosen Frozen Wave.

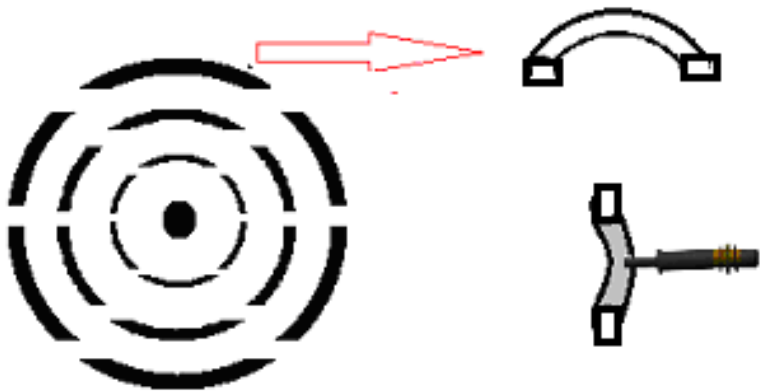

**Fig. 30** – A further possibility is discretizing the annular radiators via the introduction of rectangular (or bended) radiator elements. *Notice* that the exitation to be used for each radiating element can be obtained from the exact analytic solution which describes, for example, the chosen Frozen Wave.

Many more considerations can be found in refs.[1,2] and [13,9].

## 5.4 A brief Appendix

For possible convenience, let us here summarize (only) a standard method of ours for obtaining the Frozen Waves[17-20]. As we know, it was elsewhere extended for absorbing media (see, e.g, ref.[20,2]), in order to get FWs resisting both attenuation and diffraction.

Basically, the idea is to obtain non-diffracting beams whose desired longitudinal intensity pattern, $|F(z)|^2$, in the interval $0 \leq z \leq L$, can be chosen a priori. To obtain the desired beam we consider the following solution, given by a superposition of $2N + 1$ co-propagating and equal frequency Bessel beams of order $v$ :

$$\Psi(\rho,\phi,z,t) = e^{-i\omega t} \sum_{n=-N}^{N} A_n J_v(k_{\rho n}\rho) e^{ik_{zn}z} e^{iv\phi} \tag{10}$$

where $k_{\rho n}^2 = \omega^2/c^2 - k_{zn}^2$ and (with the choice $k_{zn} = Q + 2\pi n/L$) the parameter $Q$ is a constant obeying $0 \leq Q \pm (2\pi/L)N \leq \omega/c$.

In Eq. (1), the coefficients An are given by

$$A_n = \frac{1}{L}\int_0^L F(z) e^{-i\frac{2\pi}{L}nz} dz \tag{11}$$

where $|F(z)|^2$ is the desired longitudinal intensity pattern in the interval $0 \leq z \leq L$. This longitudinal intensity pattern can be concentrated (as we wish) over:

a) the propagation axis ($\rho = 0$). In this case $v = 0$ in eq.(10), i.e., we deal with a zero-order Bessel beam superposition. We can also choose the spot radius, $\Delta\rho_0$, of the resulting beam by making $Q = (\omega^2/c^2 - 2.4^2/\Delta\rho_0^2)^{1/2}$;

b) a cylindrical surface. In this case we deal with $v > 0$ values, and the radius $\rho_0$ of the cylindrical surface can be approximately evaluated through the equation

$$\left(\frac{d}{d\rho} J_v\left(\rho\sqrt{\omega^2/c^2 - Q^2}\right)\right)\bigg|_{\rho=\rho_0} = 0 . \tag{12}$$

# 6 – Acknowledgments

# 7 – REFERENCES


\end{center}

\end{document}